\documentclass[12pt]{interact}
\usepackage{tikz}
\usetikzlibrary{patterns,snakes,calc,decorations.pathmorphing,positioning,arrows.meta}
\usepackage[numbers,sort&compress]{natbib}
\tikzstyle{ell}=[{Latex[length=3.3,width=2.2]}-{Latex[length=3.3,width=2.2]},line width=0.3]
\tikzstyle{spring}=[line width=0.8,blue!7!black!80,snake=coil,segment amplitude=5,segment length=5,line cap=round]
\bibpunct[, ]{[}{]}{,}{n}{,}{,}
\makeatletter
\def\NAT@def@citea{\def\@citea{\NAT@separator}}
\makeatother
\begin{document}
\articletype{ARTICLE TEMPLATE}

\title{Cycling on rough roads: A model for resistance and vibration}

\author{
\name{Miles~M.\ Turner\thanks{Email: miles.turner@dcu.ie}}
\affil{School of Physical Sciences, Dublin City University, Glasnevin, Dublin 9, Ireland}
}

\maketitle

\begin{abstract}
  Minimising opposing forces is a matter of interest to most cyclists.
  These forces arise from passage through air (``drag'') and
  interaction with the road surface (``resistance'').  Recent work
  recognises that resistance forces arise not only from the
  deformation of the tyre (``rolling resistance'') but also from
  irregularities in the road surface (``roughness resistance''), which
  lead to power dissipation in the body of the rider through
  vibration.  The latter effect may also have an adverse impact on human health.
  In this work we offer a quantitative theory of roughness resistance
  and vibration that links these effects to a surface characterisation
  in terms of the International Roughness Index (IRI).  We show that
  the roughness resistance and the Vibration Dose Value (or VDV, the usual
  vibration dosage metric) can be
  expressed in terms of elementary formulae.  The roughness resistance
  depends only on the vertical stiffness of the bicycle and
  the roughness index.  Surprisingly, other
  apparently relevant parameters, such as physiological
  characteristics of the bicycle rider and other features of the
  bicycle, do not enter.   For roads of moderate roughness, roughness resistance is larger
  than rolling resistance.  For very rough roads, roughness resistance
  is larger than aerodynamic drag.  So only on roads of high quality
  (in most jurisdictions, accounting for less than 10~\% of the total)
  can roughness resistance be ignored.  Roughness resistance can be
  mitigated by reducing the vertical stiffness of the bicycle.  In common
  with other recent reports, we find that almost any cycling activity
  will breach public health guidelines relating to Vibration Dose Value.
\end{abstract}


\section{Introduction}

Bicycles and other related light vehicles encounter opposing forces
that are usually divided into ``aerodynamic drag'' and
``rolling resistance.''  Understanding the mechanisms of these forces,
with a view to their minimisation, is a central concern of the
technical literature related to bicycles, human powered vehicles more
generally, and other related kinds of vehicles such as solar cars
\cite{wilson_bicycling_2020}.  Rolling resistance is traditionally
conceived as arising from some mixture of the hysteresis losses that
occur as a rolling tyre deforms under load, and friction as parts of
the tyre slide across the surface \cite{wilson_bicycling_2020}.
Recently, another kind of resistance has attracted attention.  This is
caused by vibration associated with irregularities in the surface.
These vibrations are transmitted by the bicycle to its rider, where
they are to some extent dissipated.  The bicycle and the rider can be
imagined as a mass supported by a spring.  If the surface is not
smooth, the spring is periodically extended and compressed as the
surface is traversed, and the mass undergoes corresponding
accelerations.  If these motions are damped, then there is a net
absorption of energy, which the rider will experience as a resistance.
The coefficients involved here include physical parameters of the
bicycle, physiological features of the rider, and the character of the
road surface.  The existence of such an effect has been widely
recognised (it has been variously called ``bump
resistance,''\cite[pp.~182-187]{wilson_bicycling_2020} ``suspension
losses'' \citep{heine_all-road_2020} and ``impedance''
\cite{poertner_part_2019}) and some
experimental evidence has been gathered in informal reports \cite{poertner_part_2019,heine_all-road_2020}.  This evidence implies that this effect makes an important
contribution to the total resistance under some conditions, and that
softening the relevant spring constant will reduce the amount of
dissipation.  Cyclists do not usually speak of spring constants.  The
preferred term is ``vertical compliance'' which (of course) is
the reciprocal of the spring constant or stiffness.  An important
factor influencing the vertical compliance is the bicycle tyre.  These
considerations motivate the idea that larger tyres with lower
inflation pressures can increase the vertical compliance and
consequently reduce the additional resistance.  The influence of this
view is plainly visible in recent trends in bicycle design.

Despite the evident importance of these ideas, they have received
practically no theoretical attention.
This is
perhaps not surprising. It is not
immediately obvious how to
represent the condition of an uneven road surface in a theoretically
tractable manner, and the importance of physiological characteristics
of cyclists (which, one might think, could be highly individual) also
appears to present difficulties.  These considerations suggest that
finding a mathematically manageable model will be hard, and that even
if such a model can be found, insights of widespread validity may not
easily follow.  However, in the present work we will show that these
challenges are not as severe as they first look.  Pavement engineers
have established that as a practical matter, the condition of a road
surface can be adequately represented by a single number, a so-called
roughness index.  This is formalised in a universal way as the
International Roughness Index (IRI) \citep{sayers_guidelines_1986},
which includes a prescribed measurement procedure.  So we know how to
characterise the surface of any given road.  In addition, there is a
voluminous literature dealing with the behaviour of human bodies in
response to vibrations \cite{griffin_handbook_1996}.  Hardly any of
this addresses cycling directly, but a large amount of relevant
information is nevertheless available.  Hence there is a clear
theoretical description of rough roads and there is data to give
values to the physiological parameters that might occur in a model of
a bicycle rider.  So a suitable model indeed can be formulated.
Analysis of the model however leads to the surprising result that only
the roughness index and the vertical compliance affect the power
dissipated when a rough road is traversed, so that
\begin{equation}
  \text{Roughness resistance} \sim \frac{\left(\text{Roughness index}\right)^2}{\text{Vertical compliance}}.
  \label{eq_scaling}
\end{equation}
This resistance is associated with pavement disturbances with
wavelengths in the regime known to pavement engineers as
``roughness,'' (as we will explain below) so the term ``roughness
resistance'' seems to be appropriate.  In the view of the present
author, this is identical with ``bump resistance,'' ``suspension
losses'' and ``impedance'' (but this might be disputed). 

A related consideration is that the exposure of the cyclist to
vibrations originating at the road surface may cause injuries, or have
other adverse health effects.  Vibration dosage is a public health
concern, and there is extensive writing on the
subject \cite{griffin_handbook_1996}.  Little of this refers to
cycling, but that which does suggests that cycling activities easily
breach public health guidelines for exposure to vibrations
\cite{tarabini_whole-body_2015,duc_vibration_2016,roseiro_hand-arm_2016,doria_experimental-numerical_2021,edwards_thunder_2021}.
Surprisingly, these works in general give scant attention to
systematic road surface characterisation
(\cite{doria_experimental-numerical_2021} is an exception).  In this
work, we establish an elementary expression linking vibration dosage
to the roughness index, thus (in principle) connecting expenditure on
road maintenance to a public health issue, namely injury from exposure
to vibration.

The remainder of the paper is arranged in the following way.  In section~\ref{sec_background}
below we introduce in more detail the background topics that
have been alluded to already.  Of course, we are here presenting
only the information that is essential to the argument of the
present paper, and not a balanced overview of any of
the several large and important research areas that we refer to.
Section~\ref{sec_modelling} then develops the models that are the
central concern of this paper.  We first establish a simple model of
a bicycle travelling across a rough surface, in which rotational
motion of the bicycle about its centre of mass is neglected, and then remove this constraint.
These arguments lead to the relation Equation~\ref{eq_scaling}, in a more
specific and detailed form.  In a discussion
(section~\ref{sec_discussion}), we consider some practical
implications, by solving various more general models for bicycling to
provide comparison of the ``roughness resistance" with rolling
resistance and aerodynamic drag, and to investigate the optimal
configuration for a bicycle contending with these forces.  This
optimum, of course, depends significantly on the range of roughness
indices that will be encountered.  We also comment on the issue of
exposure to vibration.  In section~\ref{sec_conclusion} we offer a
summary and concluding reflections.


\section{Background}
\label{sec_background}

Pavement engineers distinguish between ``texture'' and ``roughness.''
Texture is a designed characteristic of a road, introduced to optimise
such things as vehicle adhesion and surface drainage.  ``Roughness,''
on the contrary is a defect of a road that should be minimized as much
as is technically and financially possible.  These properties of roads
are in practice also distinguished by wavelength.  Texture typically
is associated with wavelengths of a few centimetres or less, while
roughness occurs at longer wavelengths.  Texture may well affect the
rolling resistance of bicycles, but this is not the subject of the
present discussion.  A large amount of empirical evidence supports the
idea that road roughness has a constant power spectral density.  This
notion was codified in 1986 when the International Roughness Index (IRI) was
introduced \cite{sayers_guidelines_1986,international_organization_for_standardization_iso_2016}.
The IRI models the power spectral density of roughness
as:
\begin{equation}
  G_d(n) = G_d(n_0)\left(\frac{n_0}{n}\right)^2, \label{eq_iri}
\end{equation}
where $n$ is a wave number ({\em i.e.}, a reciprocal wavelength),
$n_0$ is a reference wave number (by convention $n_0=0.1$~m$^{-1}$),
and $G_d(n_0)$ is a coefficient characteristic of a particular
surface.  The IRI includes measurement protocols, from which
$G_d(n_0)$ can be estimated.  These protocols involve specialised
equipment, but in recent years there has been interest in presumably
less accurate but much simpler and cheaper approaches that make use of
the accelerometers that are present in most modern mobile telephones,
{\em e.g.}
\cite{zang_assessing_2018,ahmed_effects_2021,cafiso_urban_2022}.  This
development makes IRI determination much more readily available than
was once the case, which may have implications for the future elaboration
of the ideas discussed below.

\begin{figure}
  \includegraphics[width=\columnwidth]{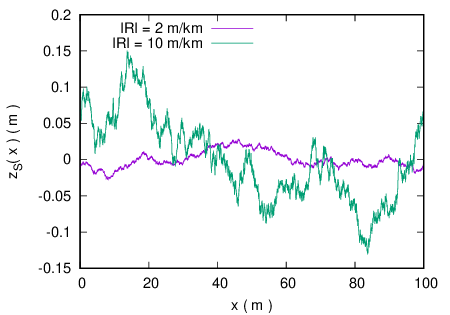}

  \caption{Vertical road surface displacement ($z_S$) as a function of
    horizontal displacement along the road surface ($x$), for two synthetic road surfaces at different International
    Roughness Indices, corresponding to a road surface of high quality
    ($IRI=2$~m/km, typical for ``airport runways and superhighways'' \cite{sayers_guidelines_1986})
    and another in poor repair ($IRI=10$~m/km, described in \cite{sayers_guidelines_1986}
    as including ``damaged pavements,'' ``maintained unpaved roads''
    and ``rough unpaved roads'' ).  These data offer some sense of the
    level of disturbance in a road surface to be expected at
    different roughness indices.\label{fig_iri_surface}}
\end{figure}

Assuming that Equation~\ref{eq_iri} applies exactly for all
wave numbers is analytically convenient (as we will see below), but
not practically particularly important, insofar as the reason for
interest in road surface roughness is usually the effect that this has on
vehicles.  Most vehicles respond appreciably only in a relatively
narrow range of wavelengths, and in this context, abundant evidence
shows that Equation~\ref{eq_iri} describes most road surfaces rather well.
An accurate description outside this range is not especially
important.  The numerical roughness index is in principle determined by $G_d(n_0)$,
but is not identical with it.  Essentially, the roughness index is
found by measuring the displacement of the surface between a finite
number of adjacent points on the road surface, and summing the
absolute values of these displacements.  An exact relationship between
the roughness index and $G_d(n_0)$ is not known, but an approximate
one is:
\begin{equation}
  IRI \approx 0.22 \sqrt{G_d(n_0)},\label{eq_iri_g_d}
\end{equation}
with $G_d(n_0)$ expressed in $10^{-6}$ m$^{3}$ and the roughness index
in m/km \citep{kropac_indicators_2007}.   Perfectly smooth roads
correspond to an IRI of 0 m/km.  High quality modern roads may have
IRI as little as 1-2 m/km, but older or less maintained roads may have
roughness indices as high as 30 m/km \cite{feighan_pavement_2004}.
Figure~\ref{fig_iri_surface} shows examples of the scale of surface
disturbances that can be expected at
two different roughness indices.  These data are synthetic
profiles generated using the method suggested
in \cite{agostinacchio_vibrations_2014}.  We will make use of
Equation~\ref{eq_iri} in formulating the model to be discussed below,
with the assumption that Equation~\ref{eq_iri_g_d} can be used to
relate $G_d(n_0)$ and the numerical value of the IRI.

\begin{figure}
  \includegraphics[width=\columnwidth]{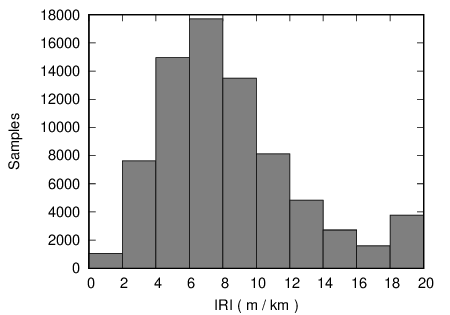}
  \caption{The distribution of the roughness index across
    secondary and tertiary roads in Ireland (after \cite{feighan_pavement_2004}).
    A ``sample'' is a 100~m section of road.  The last bin includes
    samples with $IRI > 20$~m/km.  Clearly, cyclists on such roads
    must expected to encounter roughness indices up to about 15 m/km.
    \label{fig_iri_distribution}}
\end{figure}

The range of roughness indices that cyclists might encounter in
practice will be an important consideration in the discussion below.
What might be called ``trunk'' or ``arterial'' roads, which carry
heavy traffic at high speed, generally have $IRI \lesssim 5$~m/km.
However, most cycling takes place on secondary roads \citep[Chart 26]{department_for_transport_road_2022}.  These roads
are often lightly used and little maintained, so a wide range of
roughness conditions occur.  Moreover, when the general condition of a
road is indifferent or poor, the level of roughness tends to fluctuate
rapidly.  Consequently, a cyclist has to anticipate road surface
conditions that are essentially unpredictable and rather rapidly
changing.  Figure~\ref{fig_iri_distribution} shows data drawn from a
survey of secondary and tertiary Irish roads.  On this evidence,
cyclists on these roads should be prepared for roughness indices of up
to 10-15~m/km.

A second background consideration is the response of the human body to
vibration.  There are many contexts in which human bodies are exposed
to potentially harmful vibration, and there is a correspondingly high
level of research interest \citep{griffin_handbook_1996}. The rider of
a bicycle is affected by the vibrations transmitted from the road,
which may cause discomfort or, in extreme cases, injury.  Exposure to
vibration is conventionally quantified by the Vibration Dosage Value
(VDV) \cite{griffin_handbook_1996,iso_human_2017}.  This is defined in
terms of the acceleration $a(t)$ that the body is exposed to over some
time interval $T$ as:
\begin{eqnarray}
  VDV &=& \left[\int_0^Ta^4(t)
    dt\right]^\frac{1}{4}\label{eq_vdv}\\
  &\approx& 1.4 a_\mathrm{rms}
  T^\frac{1}{4}\label{eq_evdv},
\end{eqnarray}
where the value of the second expression is known as the Estimated
Vibration Dosage Value (eVDV), and $a_\mathrm{rms}$ is the root mean
square acceleration.  (The fourth power of the acceleration
in Equation~\ref{eq_vdv}, and
the rather recherch\'e units of $VDV$ that result, are motivated by a desire
to emphasise the more extreme end of the spectrum of acceleration.)
In many jurisdictions there are statutory limits
controlling exposure to vibrations, which typically require $VDV
\lesssim 20$~m~s$^{-1.75}$.  Separately, exposure to vibrations such
that $a_\mathrm{rms} \gtrsim 10$~m~s$^{-2}$ is prohibited, or at
least deprecated.  These
criteria were not developed with cycling or any other athletic
pursuits in mind, but rather with regard to repetitive exposure in
daily life.  Their relevance to occasional sporting activity might be
doubted.  Nevertheless, where they have been applied in a sporting
context \cite{tarabini_whole-body_2015,duc_vibration_2016,roseiro_hand-arm_2016,doria_experimental-numerical_2021,edwards_thunder_2021},
they are usually found to be violated, sometimes grossly.  Whether
this is actually deleterious to cyclists appears unknown.
As an instance of what cyclists may tolerate
in practice, competitors in an approximately 30~km section of the
Paris-Roubaix bicycle race are reported to have experienced $VDV >
150$~m~s$^{-1.75}$ and $a_\mathrm{rms} \gtrsim 25$~m~s$^{-2}$
\citep{duc_vibration_2016}.

A topic directly relevant to the present work is the development of models for
bodies exposed to vibration, and these are usually in a lumped element
form involving masses, springs and damping elements
(dashpots). Although more elaborate models are available
\citep{liang_study_2006}, in this work, we have adopted a model with
two degrees of freedom (described in section~\ref{sec_modelling}
below) in the interest of analytical tractability.  The coefficients
in the model relating to the human body are chosen with reference to
the models described by \cite{muksian_frequency-dependent_1976}, which
refers to people seated without a backrest.  The values of the
parameters will be discussed below.  Also of interest, however, is the
amount of individual variation that may occur.  The available studies
essentially refer to representatives of the general population, and
suggest variations around some mean of a few tens of percent
\cite{toward_apparent_2011}.  One could suspect that practised
cyclists might prove outliers in this distribution, but we
have found no data bearing on this point.

\section{Modelling}
\label{sec_modelling}

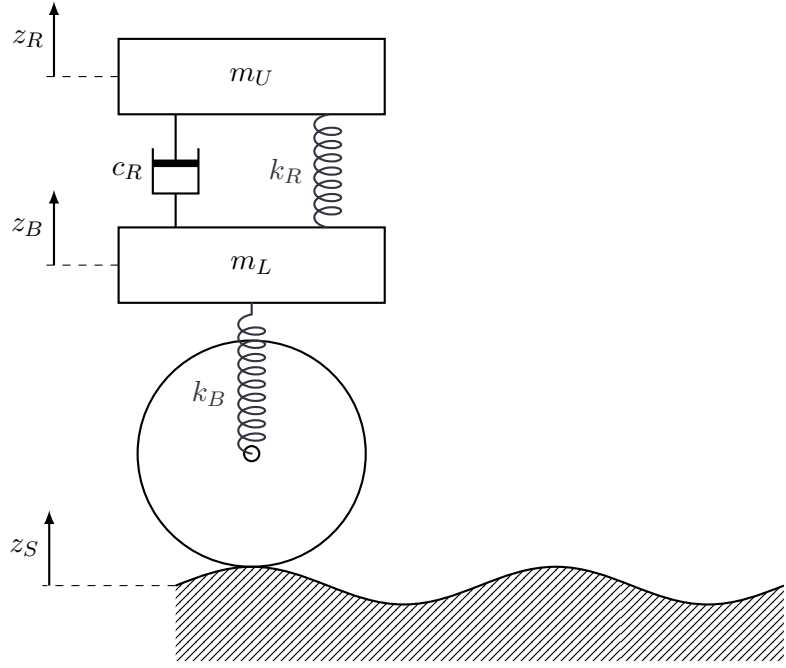
\begin{figure}
  \begin{center}
  \begin{tikzpicture}[every node/.style={outer sep=0pt},thick,
 mass/.style={draw,thick},
 ground/.style={fill,pattern=north east lines,draw=none,minimum
 width=0.75cm,minimum height=0.3cm},
 dampic/.pic={\fill[white] (-0.3,-0.3) rectangle (0.3,0.1);
 \draw (-0.3,0.3) |- (0.3,-0.3) -- (0.3,0.3);
 \draw[line width=1mm] (-0.3,0.1) -- (0.3,0.1);}]

  \draw[fill, pattern=north east lines,draw=none] (0,0) sin (1,.25) coordinate (top1) cos (2,0) sin (3,-.25) coordinate(bottom) cos (4,0) sin (5,.25) coordinate (top2) cos (6,0) sin(7,-0.25) cos(8,0) |- (0,-1)--cycle;
  \draw[thick](0,0) sin (1,.25) coordinate (top1) cos (2,0) sin (3,-.25) coordinate(bottom) cos (4,0) sin (5,.25) coordinate (top2) cos (6,0) sin(7,-0.25) cos(8,0) ;
  \draw[thick] ([yshift=15mm]top1) circle(15mm) ;
  \draw[thick] ([yshift=15mm]top1) circle(1mm) ;
  \node[mass,minimum width=3.5cm,minimum height=1cm,above=3.5cm of top1] (mB) {$m_L$} ;
  \draw[spring] ([yshift=15mm]top1) -- (mB.south) node[pos=0.4,left=2mm]{$k_B$} ;
  \node[mass,minimum width=3.5cm,minimum height=1cm,above=1.5cm of mB] (mR) {$m_U$};
  \draw ([xshift=-10mm]mB.north) -- ([xshift=-10mm]mR.south) pic[midway]{dampic} node[midway,left=3mm]{$c_R$} ;
  \draw[spring] ([xshift=10mm]mB.north) -- ([xshift=10mm]mR.south) node[pos=0.5,left=2mm]{$k_R$};
  \foreach \X in {B,R}
  {
   \draw[thin,dashed] (m\X.west) -- ++ (-1,0) coordinate[pos=0.85](aux'\X);
   \draw[-latex] (aux'\X) -- ++ (0,1) node[midway,left]{$z_\X$} ;
  }
  \draw[thin,dashed] ([yshift=-2.5mm,xshift=-10mm]top1) -- ++ (-1.75,0) coordinate[pos=0.95](auxg) ;
  \draw[-latex] (auxg) -- ++ (0,1) node[midway,left]{$z_S$} ;
  \end{tikzpicture}
  \end{center}
  \caption{A model for a bicycle crossing a rough surface, with no rotation
    of the bicycle about its centre of mass. The spring coefficient $k_B$ represents
    the combination of the bicycle frame and the tyres; $k_R$ is the stiffness of
    the riders torso (presumably mostly the spine); $m_L$ is the combined mass
    of the bicycle and the lower body of the rider ({\em i.e.} the legs and the pelvic region);
    $m_U$ is the remaining mass of the cyclist; and the damping coefficient $c_R$ is associated
    with the body of the rider.  The values assigned to these coefficients are discussed
    in Section~\ref{sec_modelling}.
    \label{fig_model_one}}
\end{figure}

We first consider the model shown in Figure~\ref{fig_model_one}.
This model imagines a cyclist with their weight carried on a saddle.
The saddle is supported by the bicycle frame, which is itself
supported by two tyres.  The supporting effect of the bicycle frame
and the tyres is represented by a stiffness constant $k_B$.  We assume that
the mass of the bicycle is negligible compared to the mass of the
rider.  The mass of the rider is divided between a lower
body mass $m_L$ and an upper body mass $m_U$.  These masses
are connected by a spring with stiffness $k_R$ with an associated
damping coefficient $c_R$, representing essentially the spine
of the rider and dissipation associated with relative motions
of the rider's upper and lower body.  We assume that
damping in the bicycle and the tyres is negligible, as we are aware of no data
to suggest otherwise.
This model is
mathematically identical to the ``quarter car model'' widely used in
the automotive literature, but the meaning attached to the physical
coefficients is different.  Reference to the literature
\cite{muksian_frequency-dependent_1976} enables us to associate
reasonable values with these coefficients, {\em viz}, $m_L = 33$~kg,
$m_U=66$~kg, $k_B=200$~kN/m, $k_R=50$~kN/m and $c_R = 1$~kN~s/m.  When
a particular value is needed (in representative plots, for example),
these values will be assumed unless there is a note to the contrary.
We will see that $k_B$ is by far the most significant of these values,
and fortunately this is a physical property of any given bicycle that
is easy to measure.  The other coefficients, of course, are not so
easily measured and may be highly personal to particular riders.  In
analysing the model shown in Figure~\ref{fig_model_one}, we assume
that linear stiffness coefficients are sufficient.  We can distinguish
(at least) two ways in which this assumption might fail.  There are
catastrophic possibilities: The bicycle could lose contact with the
surface, or a wheel rim might strike the ground, for instance.  These
are rare in practice.  A more significant incidence of nonlinearity is
the tyre response, which is definitely not linear, in general
\cite{renart_deformation_2019}.  However, small displacements around
an equilibrium can almost always be modelled by a linear response, and
we assume that this approximation can be applied.

\begin{figure}
  \begin{center}
  \includegraphics[width=\columnwidth]{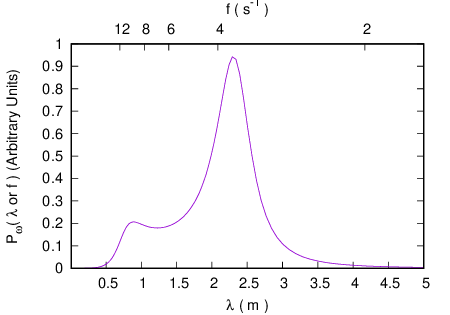}
  \caption{An example of the power dissipated in the model of
    Figure~\ref{fig_model_one} as expressed by Equation~\ref{eq_power_omega} using
    the ordinary values of the parameters discussed in the text,
    and for a bicycle travelling at 30 km/hr.  The upper axis shows temporal
    frequency and the lower axis spatial wavelength.  Of course, the relationship
    between the axes depends on the speed. Cyclists may be surprised
    to see that the dominant effect is at such large wavelengths.\label{fig_response}}
  \end{center}
\end{figure}

Inspection of Figure~\ref{fig_model_one} shows that the basic equations are:
\begin{eqnarray}
  m_L \ddot{z}_B +  c_R\left(\dot{z}_B-\dot{z}_R\right) - k_R(z_R-z_B) + k_Bz_B&=& k_Bz_S(t) \label{eq_one_z_b}\\
  m_U\ddot{z}_R + c_R\left(\dot{z}_R-\dot{z}_B\right) + k_R(z_R-z_B)  &=& 0\label{eq_one_z_r}.
\end{eqnarray}
which are more conveniently written using  $z_C = z_R-z_B$:
\begin{eqnarray}
  m_L \ddot{z}_B -  c_R\dot{z}_C - k_Rz_C + k_Bz_B&=& k_Bz_S(t)\\
  \mu \ddot{z}_C + c_R\dot{z}_C + k_Rz_C
    - \frac{k_B \mu }{m_L}z_B&=& - \frac{k_B \mu}{m_L}z_S(t).
\end{eqnarray}
where $\mu = m_Um_L/(m_U+m_L)$.
We assume all time variation is harmonic with angular frequency $\omega$, such that (for example) $z_C(t) = \tilde{z}_C\exp\left(-\omega t\right)$, where
$\tilde{z}_C$ is an (in general) complex amplitude with complex conjugate denoted by $\tilde{z}^*_C$.
Then we can show:
\begin{eqnarray}
\frac{\tilde{z}_C}{\tilde{z}_S} &=&
\frac{\omega^2\left(\frac{k_B}{m_L}\right)}{\omega^4 + \frac{c_R}{ \mu} i\omega^3
  - \left[\frac{k_R}{\mu}+\frac{k_B}{m_L}\right]\omega^2
 - \left(\frac{c_Rk_B}{m_Um_L}\right)i\omega + \frac{k_Bk_R}{m_Lm_U}}.
\end{eqnarray}
Of concern is the power dissipated in the dashpot, which is given by
\begin{eqnarray}
  P_\omega &=& \frac{1}{2} c_R \omega^2 \tilde{z}_C \tilde{z}_C^*\\
  &=&
  \frac{1}{2}\frac{c_R\left(\frac{k_B}{m_L}\right)^2\omega^6\tilde{z}_S\tilde{z}^*_S}
       {\left[\omega^4 - \left(\frac{k_R}{\mu}+\frac{k_B}{m_L}\right)\omega^2 +  \frac{k_Bk_R}{m_Lm_U}\right]^2
         +\left[\left(\frac{c_R}{\mu}\right)\omega^3-\frac{c_Rk_B}{m_Um_L}\omega\right]^2}.\label{eq_power_omega}
\end{eqnarray}
An example of this response is shown in Figure~\ref{fig_response}.
This is (unsurprisingly) bimodal and shows that for these parameters
(the ones discussed above), the maximum effect is at temporal
frequencies of about 2-10~Hz and spatial wavelengths of 0.5-4~m.
These spatial wavelengths of course depend on the speed of the bicycle
which in this example is assumed to be 30~km/hr.  Lower speeds will
shift the response to shorter wavelengths, and {\em vice versa}.  At
normal cycling speeds, therefore, the response is caused by
``roughness'' and not ``texture'' in the sense of these terms
discussed above.

We now ask:  What effect is produced when the disturbance $\tilde{z}_S$ is due
to surface roughness characterised by Equation~\ref{eq_iri}?  If we consider a region
around the wave number $n$ with infinitesimal size $dn$, then the effective
amplitude of the disturbance in this region is \cite{agostinacchio_vibrations_2014}:
\begin{equation}
  \tilde{z}_S\tilde{z}_S^* = 2 G_d(n) dn.
\end{equation}
Then, noting that $\omega = 2\pi n v$, for a bicycle travelling at
speed $v$, and introducing the constant $\omega_0 = \left(k_B
c_R/m_L^2\right)^\frac{1}{3}$ and the dimensionless variable of
integration $s=\omega/\omega_0$, we can write the wave number
integrated power dissipation $P$ as
\begin{eqnarray}
  P &=& 2\pi v k_B n_0^2 G_d(n_0) I_0
\end{eqnarray}
where
\begin{eqnarray}
  I_0 &=& \int_0^\infty \frac{s^4ds}
  {\left[s^4 - \left(\frac{k_R}{\mu}+\frac{k_B}{m_L}\right)\frac{s^2}{\omega_0^2} + \frac{k_Bk_R}{\omega_0^4m_Lm_U}\right]^2
    +\left[\frac{c_R}{\mu\omega_0} s^3-\frac{c_Rk_B}{m_Um_L\omega_0^3} s\right]^2}\label{eq_roughness_integral}\\
  &=& \frac{\pi}{2},
\end{eqnarray}
where $I_0$ is an instance of integral 3.112 of \cite[p. 253]{gradshteyn_table_2007}, which can be evaluated
using the formulae given therein.  Hence:
\begin{equation}
  P = \pi^2 v k_B n_0^2 G_d(n_0) = F_\mathrm{ro} v.\label{eq_roughness_resistance}
  \end{equation}
Equation~\ref{eq_scaling} follows from this result, combined with Equation~\ref{eq_iri_g_d}.  This
outcome appears surprising, but similar
results have been found in other dissipative systems excited by random processes
\citep{clough_dynamics_1993,pippard_physics_1989,langley_can_1997,smith_power_2011,clark_power_2012,langley_general_2014},
some of them considerably more complex than the one under
consideration here. The simplicity of the result, of course, depends on the assumption
that Equation~\ref{eq_iri} is valid for all wave numbers.  This cannot
be exactly true.  Even if the road surface is accurately described by
Equation~\ref{eq_iri}, the response of the bicycle must have a high
frequency limit associated with the finite length of the tyre contact
patch.  However, Figure~\ref{fig_response} shows that dominant part of
the response is at wavelengths appreciably longer than the size of the
contact patch, so this does not appear to be an important limitation
of the present model.

The Vibrational Dose Value is dependent on the root mean square acceleration,
which can be calculated in a similar way.
By considering $\ddot{z}_C$, we can show that
\begin{equation}
a_\mathrm{rms} = \pi\sqrt{\frac{v G_d(n_0)n_0^2 k_B^2}{2 m_L c_R}}\label{eq_a_rms}.
\end{equation}
Alternatively, making use of $\ddot{z}_B$ leads to:
\begin{equation}
a_\mathrm{rms} = \pi\sqrt{\frac{v G_d(n_0)n_0^2}{2}\left[
    \frac{c_Rk_B}{m_U^2}
    + \frac{\left(m_Uk_B - m_L k_R\right)^2}{c_Rm_U^2m_L}\label{eq_a_rms_b}
    + \frac{k_R^2}{c_Rm_U}\right]}.
\end{equation}
Either of these can be used in conjunction with Equation~\ref{eq_evdv}
to estimate the Vibration Dose Value, but only
Equation~\ref{eq_a_rms_b} conforms with the prescribed standard method
for VDV calculation \cite{iso_human_2017}, which requires
$a_\mathrm{rms}$ to be evaluated at the point of entry into the body
(the saddle, in this case, so the acceleration at $z_B$).  The
standard also requires the use of a frequency filter
\cite{iso_human_2017,goncalves_optimization_2003}.  However, the frequency
filter has insignificant effect over the range suggested by
Figure~\ref{fig_response}.  Perhaps more practically important than
these details is the difficulty of precisely defining the values of
the parameters.

\begin{figure}
  \begin{center}
  \begin{tikzpicture}[every node/.style={outer sep=0pt},thick,
 mass/.style={draw,thick},
 ground/.style={fill,pattern=north east lines,draw=none,minimum
 width=0.75cm,minimum height=0.3cm},
 dampic/.pic={\fill[white] (-0.3,-0.3) rectangle (0.3,0.1);
 \draw (-0.3,0.3) |- (0.3,-0.3) -- (0.3,0.3);
 \draw[line width=1mm] (-0.3,0.1) -- (0.3,0.1);}]

  \draw[fill, pattern=north east lines,draw=none] (0,0) sin (1,.25) coordinate (top1) cos (2,0) sin (3,-.25) coordinate(bottom) cos (4,0) sin (5,.25) coordinate (top2) cos (6,0) sin(7,-0.25) cos(8,0) |- (0,-1)--cycle;
  \draw[thick](0,0) sin (1,.25) coordinate (top1) cos (2,0) sin (3,-.25) coordinate(bottom) cos (4,0) sin (5,.25) coordinate (top2) cos (6,0) sin(7,-0.25) cos(8,0) ;
  \draw[thick] ([yshift=15mm]top1) circle(15mm) ;
  \draw[thick] ([yshift=15mm]top1) circle(1mm) ;
  \draw[thick] ([yshift=15mm]top2) circle(15mm) ;
  \draw[thick] ([yshift=15mm]top2) circle(1mm) ;
  \node[mass,minimum width=7cm,minimum height=1.5cm,above=4.5cm of bottom,label={[xshift=4mm,yshift=-7.5mm]{$m_L$}}] (m1) {} ;
  \draw[spring] ([yshift=15mm]top1) -- ([xshift=-2cm] m1.south) node[pos=0.2,left=2mm]{$k_{WB}$} ;
  \draw[spring] ([yshift=15mm]top2) -- ([xshift=2cm] m1.south) node[pos=0.2,left=2mm]{$k_{WF}$} ;
  \node[mass,minimum width=3.5cm,minimum height=1.5cm,above=1.5cm of m1,label={[xshift=4mm,yshift=-7.5mm]{$m_U$}}] (m2) {};
  \draw ([xshift=-10mm]m1.north) -- ([xshift=-10mm]m2.south) pic[midway]{dampic} node[midway,left=3mm]{$c_R$} ;
  \draw[spring] ([xshift=10mm]m1.north) -- ([xshift=10mm]m2.south) node[pos=0.5,left=2mm]{$k_R$};
  \draw[ell] ([xshift=-2cm,yshift=0.25cm] m1.south) -- ([xshift=2cm,yshift=0.25cm] m1.south) node[midway,fill=white,inner sep=0]{$L_W$} ;
  \draw[ell] ([xshift=-2cm,yshift=0.5cm] m1.south) -- ([xshift=0cm,yshift=0.5cm] m1.south) node[midway,fill=white,inner sep=0]{$x_M$} ;
  \draw[thick] ([yshift=-7.5mm]m1.north) circle(1mm) ;
  \draw[thick] ([yshift=-7.5mm]m2.north) circle(1mm) ;
  \end{tikzpicture}
  \end{center}
  \caption{A model for a bicycle that is able to rotate about its centre of
    mass as it passes across the surface.\label{fig_model_two}}
\end{figure}
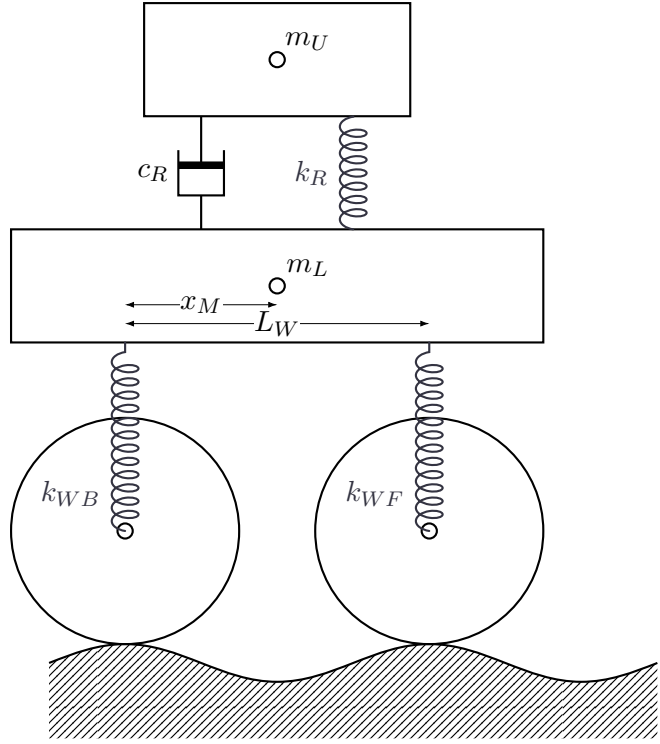

We consider now a model with two wheels, shown in
Figure~\ref{fig_model_two}.  (This incidentally is not the same as the
``half car model'' often used for road vehicles.)  We assume that both
wheels follow the same path across the road surface, so that the
disturbances at each wheel differ only by a time delay, $\Delta t$.
The motion in this case is a superposition of translations of the
centre of mass and rotations about the centre of mass.  Relative to
the model discussed above, the extra effect that is introduced is the
possibility that the vehicle can tilt as the wheels pass over the road
surface. The nature of this effect varies with the wavelength of
disturbances in the road surface.  If the wheelbase is an integral
multiple of the wavelength, there is no effect, because both wheels
rise and fall in phase with each other.  If, however, the wheelbase
is a half integral multiple of the wavelength, and the centre of mass
is midway between the wheels, the motion is a pure rotation around the
centre of mass, and there is no vertical displacement of the centre of
mass.  One might therefore guess that the effect will be to reduce the
resistance by a factor of two, and this is not far from the truth, as
the detailed argument shows.

\begin{figure}
  \includegraphics[width=\columnwidth]{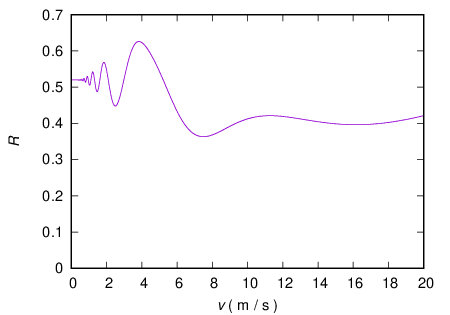}
  \caption{The rocking factor $R$ found by integrating Equation~\ref{eq_rocking}
    over the model response.  In detail this curve depends on all
    of the model parameters.  This example was computed using the
    parameters discussed in the text.\label{fig_rocking}}
\end{figure}

Consider a frame of reference in which the bicycle is stationary.  In this frame
of reference, the rear wheel contacts the ground at $x_{WB}$ and the front wheel
at $x_{WF}$ so that the wheel base $L_W=x_{WF}-x_{WB}$.  The bicycle has a centre of
mass at $x_M$, and naturally $x_{WB} \le x_M \le x_{WF}$.  Road surface roughness
causes vertical displacements at each wheel, and these displacements are denoted by $z_{WB}$
and $z_{WF}$.  By convention, these quantities are taken to be zero when the bicycle is
at rest on a flat surface.  When $z_{WB}$ and $z_{WF}$ are not zero, the centre
of mass will be displaced vertically by an amount $z_M$ and a rotation
about the centre of mass by an angle $\theta$ will occur, so that:
\begin{eqnarray}
  z_M &=& \left(1-\xi_M\right) z_{WB} + \xi_M z_{WF} \\
  \theta &=& \frac{z_{WF}-z_{WB}}{L_W}
\end{eqnarray}
where $\xi_M = x_M/L_W$ and we assume $\sin\theta \approx \theta$.  The passage
of the bicycle over the road surface produces a displacement of the surface at
$z_{WB}$ given by $z_S(t)$ and at $z_{WF}$ given by $z_S(t+\Delta t)$, where the time
constant $\Delta t$ is $L_W/v$ when the bicycle travels at speed $v$.  A force constant is
associated with each of these displacments, such that the vertical force at
the center of mass is
\begin{eqnarray}
  F_M &=& F_{WB} + F_{WF}\\
  &=& -k_{WB}\left[z_{WB} - z_S(t)\right] - k_{WF} \left[z_{WF} - z_S(t+t_0)\right]\\
  &=& -\left(k_{WB}+k_{WF}\right)z_M + L_W\left[\xi_M k_{WB} - (1-\xi_M)k_{WF}\right]\theta\nonumber\\
  && + k_{WB} z_S(t) + k_{WF} z_s(t+\Delta t)
\end{eqnarray}
and the torque around the centre of mass is
\begin{eqnarray}
  T_M &=& L_W\left[\left(1-\xi_M\right) F_{WF} - \xi_M F_{WB}\right]\\
  &=& -\left\{L_W\left[(1-\xi_M)k_{WF} - \xi_M k_{WB}\right]z_M + L_W^2\xi_M\left[\xi_c k_{WB} + (1-\xi_M)k_{WF}\right]\theta\right\}\nonumber\\
  && -L_W\left[\xi_M k_{WB} z_S(t) - (1-\xi_M)k_{WF}z_S(t+t_0)\right].
\end{eqnarray}
We ensure that the bicycle is horizontal at equilibrium by assuming
\begin{eqnarray}
  k_{WB} &=& \left(1-\xi_M\right) k_B \label{eq_k_wb}\\
  k_{WF} &=& \xi_M k_B \label{eq_k_wf}
\end{eqnarray}
so that
\begin{eqnarray}
  F_M &=& -k_B z_c + k_B\left[(1-\xi_M)z_S(t) + \xi_M z_S(t+\Delta t)\right]\label{eq_f_m}\\
  T_M &=& -2k_BL_W^2\xi_M^2(1-\xi_M)\theta - k_B L_W(1-\xi_M)\xi_M\left[z_S(t) - z_S(t+\Delta t)\right]. 
\end{eqnarray}
Equations~\ref{eq_k_wb} and \ref{eq_k_wf} uncouple the vertical
displacement from the angular displacement.  Moreover, in the
approximation $\cos\theta \approx 1$ (for small angular
displacements), movements of the masses parallel to the $x$ axis can
be neglected.  Hence we can identify $z_M$ with $z_B$ of the previous
model, and write the equations of motion for the displacements along
the $z$ axis as ({\em c.f.} Equations~\ref{eq_one_z_b}-\ref{eq_one_z_r}):
\begin{eqnarray}
  m_L \ddot{z}_B +  c_R\dot{z}_C - k_Rz_C + k_Bz_B
  &=& k_B\left[(1-\xi_M)z_S(t) + \xi_M z_S(t+\Delta t)\right]\\
  m_U\ddot{z}_R + c_R\dot{z}_C + k_Rz_C  &=& 0,
\end{eqnarray}
and we see that the effect of introducing the rocking motion is only to modify the forcing
term.  In the frequency domain, the time delay $\Delta t$ becomes a phase difference
\begin{equation}
  \phi = \frac{2\pi L_W}{\lambda} = \frac{\omega L_W}{v}\label{eq_phase}
  \end{equation}
where $\lambda$ is a spatial wavelength and $v$ is the speed of the bicycle.  Consequently, the effect
of introducing the rocking motion is to modify the driving term in the model, such that:
\begin{equation}
  {{\tilde{z'}}_S}^2 = \left[\left(1-\xi_M\right)^2 + 2\left(1-\xi_M\right)\xi_M\cos\phi + \xi_M^2\right]\tilde{z}_S^2.
\end{equation}
So the required change is to insert the rocking factor
\begin{equation}
  R_\omega = \frac{ {{\tilde{z'}}_S}^2}{\tilde{z}_S^2}
  = \left(1-\xi_M\right)^2 + 2\left(1-\xi_M\right)\xi_M\cos\phi + \xi_M^2\label{eq_rocking}.
  \end{equation}
into the integral in Equation~\ref{eq_roughness_integral}.  Unfortunately
we have found no exact treatment of the integral that arises when the
rocking factor is included.  The value of this integral is a function
of the speed $v$ of the bicycle because of the phase factor given by
Equation~\ref{eq_phase}.  Values of the rocking factor as a function of
speed are shown in Figure~\ref{fig_rocking}, where the integral has been
computed by a quadrature. The oscillations arise because the
cosinusoidal term in Equation~\ref{eq_rocking} is convoluted with a
response such as the one shown in Figure~\ref{fig_response}.  Evidently,
this leads to the remarkable situation that the roughness resistance
and possibly the total resistance are not always monotonically
increasing functions of speed.  This effect can be more extreme for
parameters different from the base values discussed above.  For
example, reducing the damping coefficient can lead to more pronounced
oscillations.  This might conceivably be an exploitable effect in
situations where a bicycle travels at nearly constant speed for long
intervals, such as in a time trial on level ground.  However, more
generally this seems likely to be an exotic curiosity.  For the
purposes of further discussion, we assume that for a typical cyclist
travelling at variable speed, the oscillatory effects average out and
so we approximate the rocking factor as
\begin{equation}
  R \approx \left(1-\xi_M\right)^2 + \xi_M^2\label{eq_rocking_approx}.
\end{equation}
A typical value of $\xi_M$ is about 0.4 \cite{yizhaq_new_2010}, so
that $R \approx 0.52$.  This correction applied to
Equation~\ref{eq_roughness_resistance} is the roughness resistance that we
will employ in further discussion.  Similarly, the rocking factor
can be inserted into Equation~\ref{eq_a_rms} or \ref{eq_a_rms_b} to estimate
the Vibration Dose Value.

\section{Discussion}
\label{sec_discussion}

Understanding the significance of the results obtained above requires
that we link the roughness resistance with parameters of transparent
practical significance, and consider the relationship of the roughness
resistance, the rolling resistance and the aerodynamic drag.  Rolling
resistance is not, in general, quantitatively well understood
\cite{wilson_bicycling_2020}, meaning there is no theory available to
calculate the rolling resistance from first principles.  There are,
however, extensive measurements of the coefficient of rolling resistance
$c_\mathrm{rr}$
(see, {\em e.g.}, \citep{biermann_bicycle_2023}).  The most important parameters are the
tyre inflation pressure $P$ and the tyre radius $R_{T0}$.  A convenient
form is
\begin{equation}
  c_\mathrm{rr} = c_\mathrm{rr}^{(0)}
  \left[\frac{P^{(0)}R^{(0)}_{T0}}{P R_{T0}}\right]^\frac{1}{2},
  \label{eq_crr_scal}
\end{equation}
where the coefficient $c_\mathrm{rr}^{(0)}$ is inferred from experiments, while
$P^{(0)}$ and $R^{(0)}_{T0}$ are reference values.

We have seen that the stiffness of the bicycle ($k_B$) is one of
the most important factors affecting the roughness resistance, and this
stiffness is strongly influenced by the tyre stiffness, $k_T$.  Hence
it is important to have a quantitative understanding of the
relationship between the physical parameters characterising a bicycle
tyre, and the stiffness of the same tyre.  A convenient model is described
in \citep{renart_deformation_2019}, in which the bicycle wheel and
tyre is characterised by the tyre radius under no load, $R_{T0}$, the
rim width, $W_L$, the wheel radius, $R_W$, and the inflation pressure,
$P$.  (Where a numerical value is required later, we assume
$W_L=2R_{T0}$.)  This model shows that if the tyre deflection from the
no load condition is $d$, then the force exerted is
\begin{equation}
  F_T \approx \alpha\beta P d^\frac{3}{2},
\end{equation}
where $\alpha=1.259$ is a dimensionless numerical constant and
$\beta$ is a geometrical factor involving complex sub-formulae that we will not quote.
If the tyre bears a load $M_L$ then the equilibrium deflection is
\begin{equation}
  d = \left(\frac{M_L g}{\alpha\beta P}\right)^\frac{2}{3}.
\end{equation}
We assume that normal cycling is characterised by small changes in
deflection near this equilibrium point, so that the linear force constant
is
\begin{equation}
  k_T = \left.\frac{dF_T}{dz}\right|_{z=d} = \frac{3}{2}\left(\alpha\beta P\right)^\frac{2}{3}
  \left(M_L g\right)^\frac{1}{3}.\label{eq_k_t}
\end{equation}
We will be interested in establishing the smallest force constant that
can be achieved.  For fixed geometrical parameters, this requires
choosing the smallest permissible tyre pressure.  Standards
developed by the European Tyre and Rim Technical Organisation
allow a 30\% maximum tyre deflection at equilibrium load
\cite[p.~M9]{european_tyre_and_rim_technical_organisation_standards_2003},
relative to the unloaded tyre height $2R_{T0}$ and this condition
implies a minimum tyre pressure given by
\begin{equation}
  P_\mathrm{min} = \frac{M_L
    g}{\alpha\beta\left(0.6R_{T0}\right)^\frac{3}{2}}.\label{eq_p_min}
\end{equation}
and so the minimum force constant is
\begin{equation}
  k_{T,\mathrm{min}} = \frac{M_Lg}{0.6R_{T0}},\label{eq_k_t_min}
\end{equation}
surprisingly independent of the rim width $W_L$, although the
minimum pressure is not. There is also a maximum spring constant,
associated with the maximum type pressure.  What determines this?
The tension in a tyre casing of thickness $h$ is
\citep{renart_deformation_2019}:
\begin{equation}
  \sigma = \frac{R_{T0}P}{h}.\label{eq_p_max}
\end{equation}
For a given tyre casing construction (with $h$ fixed, and some
$\sigma_\mathrm{max}$, above which the tyre will fail), there is
therefore a maximum inflation pressure inversely proportional to the
radius, such that $R_{T0} P_\mathrm{max}$ is constant.  Under this
constraint, the rolling resistance when $P=P_\mathrm{max}$ is
independent of $R_{T0}$.  That this holds exactly here is contingent on
the particular forms of Equation~\ref{eq_crr_scal} and the
tyre model.  However, for any reasonable alternatives to these, the
result will hold at least approximately.

\begin{figure}
\begin{center}
  \includegraphics[width=0.9\columnwidth]{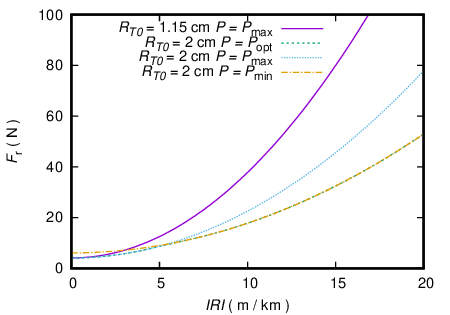}

  (a)
  
  \includegraphics[width=0.9\columnwidth]{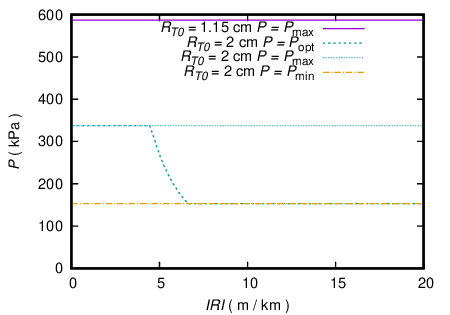}

  (b)
  
  \caption{(a) The total resistance force $F_\mathrm{r}=F_\mathrm{rr} + F_\mathrm{ro}$ as a function of the roughness
    index for bicycles with stiff frames ($k_F=1$~MN/m).  Data are shown for
    a traditional high performance tyre with $2R_{T0}=23$~mm inflated to the
    maximum permissible pressure, and for a modern ``all road'' tyre ($2R_{T0}=40$~mm)
    with three different pressure choices:  The maximum allowed ($P_\mathrm{max}$),
    the minimum allowed ($P_\mathrm{min}$), and the pressure that minimizes
    the resistance force at a given value of $IRI$ ($P_\mathrm{opt}$).

    (b)  The tyre pressures corresponding to the data shown in (a), showing that
    the optimal solution is $P_\mathrm{max}$ for $IRI \lesssim 5$~m/km, $P_\mathrm{min}$
    for  $IRI \gtrsim 10$~m/km, with a transitional regime in between.  (Recall
    that 100~kPa~$\approx$~1~Bar~$\approx~15$~psi.)
    \label{fig_resistance}}
  \end{center}
\end{figure}

\begin{figure}
\begin{center}
  \includegraphics[width=0.9\columnwidth]{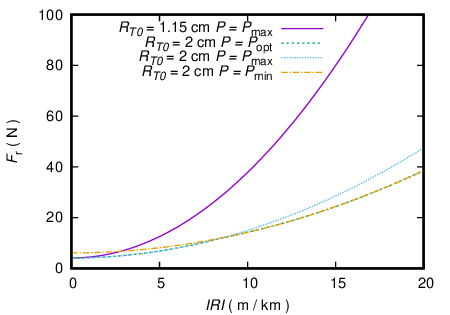}

  (a)
  
  \includegraphics[width=0.9\columnwidth]{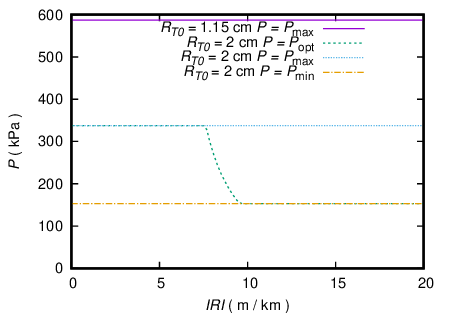}

  (b)
  
  \caption{  (a)  As in Figure~\ref{fig_resistance}, with the difference that the second
  bicycle with $2R_{T0}=40$~mm has a frame that is appreciably less stiff, with $k_F=200$~kN/m
  rather than $k_F=1$~MN/m.  Evidently, this change substantially reduces
  the resistance force.

  (b) As in Figure~\ref{fig_resistance}, showing that the transition between the
  high and low pressure optimal solutions is postponed to $IRI > 10$~m/km
  by reducing the frame stiffness.
  \label{fig_resistance_frame}}
  \end{center}
\end{figure}

These preliminaries enable us to address the practical implications of
Equation~\ref{eq_roughness_resistance}.  Let us first consider a
traditional high performance bicycle, with a stiff frame
($k_F=1$~MN/m) and narrow tyres ($2R_{T0}=23$~mm).  The rolling
resistance parameters in Equation~\ref{eq_crr_scal} are chosen using data
from \citep{biermann_bicycle_2023} to be representative of
contemporary high performance tyres.  The curve in the upper panel of
Figure~\ref{fig_resistance} labelled $R_{T0}=1.15$~cm and
$P=P_\mathrm{max}$ shows the resistance force encountered by this
bicycle as a function of the roughness index.  An initial point is
that the presence of roughness significantly increases the total
resistance on practically any road.  Even at the upper limit of well
maintained roads ($IRI=5$~m/km), the resistance is more than doubled,
and at $IRI=10$~m/km, the resistance is six times larger.  This
implies that roughness resistance is a matter of concern for almost
all cyclists, whether (for examples) engaged in competition on
well-maintained roads, or audax events during which poorer surfaces
predominate.  What can be done to minimise the roughness effect?
Clearly, because of Equation~\ref{eq_roughness_resistance},
only one recourse is available, and that is to reduce the
stiffness of the bicycle.  In general, and supposing $k_T$ is
specified for one tyre:
\begin{equation}
  k_B \approx \frac{2 k_T k_F}{2 k_T + k_F}.\label{eq_bicycle_stiffness}
\end{equation}
So one can reduce $k_B$ by reducing either $k_F$ or $k_T$ or both.
Let us consider these options separately, and address tyre stiffness
first.  Equation~\ref{eq_k_t} shows that the simplest measure, which
is always available, is to reduce the tyre pressure.  If this is
insufficient, then the tyre radius can be increased as well, and the
minimum stiffness available is then given by
Equation~\ref{eq_k_t_min}.  Equation~\ref{eq_k_t_min} suggests that
the stiffness can be reduced without bound by increasing $R_{T0}$,
but this course of action obviously has practical limits.  As an example,
we consider $2R_{T0}=40$~mm, but the choice here does not affect the
conclusions that will be drawn.  Also plotted in
Figure~\ref{fig_resistance} are the
resistance forces for this larger tyre for three pressure
cases:  $P_\mathrm{max}$, $P_\mathrm{min}$ and the pressure $P_\mathrm{opt}$
that minimises the total resistance.  On smoother roads
($IRI \lesssim 5$~m/km), the
optimal choice is $P_\mathrm{opt}=P_\mathrm{max}$ and in this regime the
total resistance is indistinguishable from the
narrower tyre considered above.  On rougher roads ($IRI \gtrsim 10$~m/km)
the optimal choice is $P_\mathrm{opt}=P_\mathrm{min}$ and here the total resistance
is much less than the narrower tyre, by a factor of between 1.5 and 2
in the range shown.  In the transitional regime $5 \lesssim IRI \lesssim 10$~m/km,
there is a corresponding transition in the optimal pressure, which
is shown in the lower panel of Figure~\ref{fig_resistance}.  These data show
that the wider tyre has an optimal resistance that is never higher
than the narrower tyre, but significantly lower when $IRI \gtrsim 5$~m/km.
The difficulty here, of course, is that cyclists cannot usually
predict in advance what range of roughness index they will encounter,
and nor (in most cases) can they rapidly adapt their tyre pressure to changing road conditions.
Let us now turn to the other possibility, namely reducing the frame
stiffness $k_F$.  Reports in the cycling press suggest that
the stiffness of frames has fallen in recent years.  Typical
now seems to be $100 \lesssim k_F \lesssim 300$~kN/m, with
some instances of $k_F \lesssim 100$~kN/m.  Figure~\ref{fig_resistance_frame}
shows the same calculations as Figure~\ref{fig_resistance} with the exception
that $k_F=200$ kN/m.  Clearly, the regime in which the
narrower and wider tyres are indistinguishable has extended
to $IRI \lesssim 10$~m/km and the rough regime is postponed
to $IRI \gtrsim 15$~m/km.  Of course, frames with even lower $k_F$ (which exist)
will move these regimes to even larger $IRI$.  Consequently, reduced
frame stiffness achieves much the same outcome in respect of
resistance as reduced tyre stiffness, but without the practical complication
of varying the tyre pressure, because the optimal pressure for all but
the most extreme road conditions is $P_\mathrm{max}$.  The clear implication
is that a highly compliant ({\em i.e.} low stiffness) frame is
the most practical way of reducing the resistance experienced on
rough roads.

\begin{figure}
  \begin{center}
    
    \includegraphics[width=0.9\columnwidth]{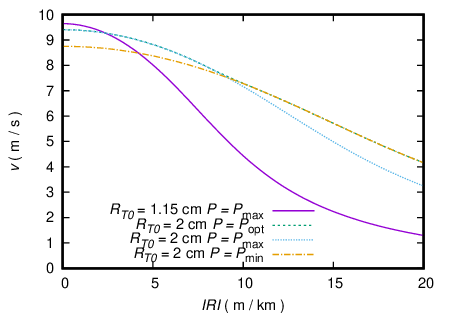}

    (a)

    \includegraphics[width=0.9\columnwidth]{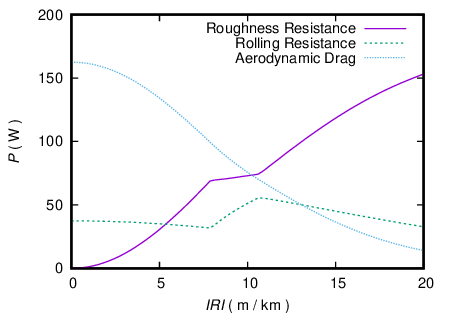}

    (b)
  \caption{(a) Bicycle speeds obtained by solving the power balance
    Equation~\ref{eq_power_balance} including aerodynamic drag, roughness resistance,
    and rolling resistance, for a cyclist with $P_C=200$~W.  The bicycles are
    the same as in Figure~\ref{fig_resistance_frame}, but the bicycle with larger
    tyres is assumed to have increased drag area as discussed in the text.
    This additional drag reduces speed for insufficient compensation when
    $IRI \lesssim 5$~m/km.  On roads with $IRI \gtrsim 5$, roughness resistance
    is more important and the drag area less so.  Reduced frame stiffness
    is also important, as in Figure~\ref{fig_resistance_frame}.

    (b) The total power divided between roughness resistance, rolling
    resistance and aero dynamic drag for total power $P_C=200 W$, for the
    case in (a) with $2R_{T0}=4$~cm, $P =P_\mathrm{opt}$ and $k_F=200$ kN/m.

    \label{fig_aerodrag}}

  \end{center}
\end{figure}

\begin{figure}
  \includegraphics[width=0.9\columnwidth]{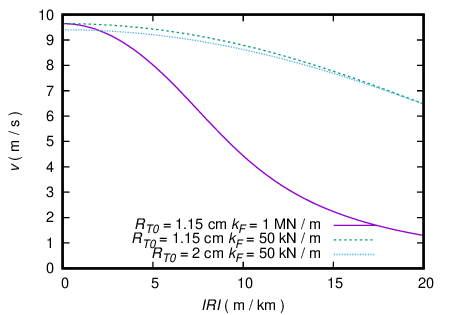}
  \caption{These data are similar to Figure~\ref{fig_aerodrag}, but
    compare bicycles with different frame stiffness.  In particular,
    considered here is the case where $k_F=50$~kN/m, approximately
    the lowest value that seems presently practical.  In all these
    cases $P=P_\mathrm{max}$.  Under these conditions, the
    bicycle stiffness is dominated by the frame stiffness (see Equation~\ref{eq_bicycle_stiffness}),
    and tyre stiffness has almost no influence.
    \label{fig_aerodrag_soft}}
\end{figure}

Aerodynamic drag
\citep{crouch_riding_2017,malizia_bicycle_2020,wilson_bicycling_2020}
is generally proportional to $v^2$ and is characterised by the drag
area $C_DA$, where $C_D$ is a dimensionless drag coefficient and $A$
is a relevant area.  This is usually a larger force than the
resistance we have been examining, and it is affected by the choice of
tyre size, because larger tyres increase the aerodynamic
drag. Typically the drag area for a cyclist varies from around
0.2~m$^{2}$ to 0.4~m$^{2}$
\citep{crouch_riding_2017,malizia_bicycle_2020}.  For the purpose of
this discussion, we select the middling value $C_DA=0.3$~m$^{2}$.
With reference to wind tunnel data \citep{crane_drag_2018}, we assume
that a tyre wider than the lowest value we have considered
($2R_{T0}=23$~mm) adds to the drag area at a rate of 0.015~m$^{2}$ per
cm of additional tyre diameter, which means that when $2R_{T0}=4$~cm,
the drag area is increased by about 10~\%.  We evaluate the effect of
this additional drag by solving a power balance equation:
\begin{equation}
  P_C= \frac{1}{2}\rho C_DA v^3 + \left(c_\mathrm{rr} M g  + F_\mathrm{ro}\right)v^2,\label{eq_power_balance}
\end{equation}
where $P_C$ is the power available from the cyclist and $\rho$ is the density
of air (assumed to be 1.2~kg m$^{-3}$).  The results of
this calculation are shown in Figure~\ref{fig_aerodrag}, which compares
the traditional high performance bicycle discussed above
($2R_{T0}=23$~mm and $k_F=1$~MN/m) with an otherwise similar machine
with $2R_{T0}=4$~cm and $k_F=200$~kN/m (as in
Figure~\ref{fig_resistance_frame}).  These data show that for
$IRI \lesssim 5$~m/km, the additional aerodynamic drag
caused by the larger tyre is appreciable.  In competitions
such as time trials, the speed difference shown here would
be important.  In Figure~\ref{fig_aerodrag_soft}, we
consider the case of a bicycle frame with exceptionally
low stiffness, $k_F=50$~kN/m.  In this case, the bicycle
stiffness given by Equation~\ref{eq_bicycle_stiffness} is dominated by
the frame stiffness, and tyre stiffness is barely a factor.
Consequently, the additional aerodynamic drag of a larger
tyre is a disadvantage at all values of IRI.  But this is
unlikely to be the only consideration in practice.
Attempting to ride a narrow tyre on a surface with $IRI > 10$~m/km
could be a highly unsatisfactory experience, that could well
mitigate in favour of a larger tyre despite the
apparent aerodynamic limitations.  In addition, the question
of what degree of stiffness is desirable in a bicycle is complex:
There are considerations other than minimising
roughness resistance.

\begin{figure}
  \begin{center}
    \includegraphics[width=0.9\columnwidth]{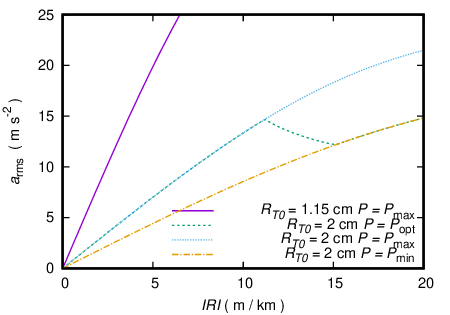}

    (a)

    \includegraphics[width=0.9\columnwidth]{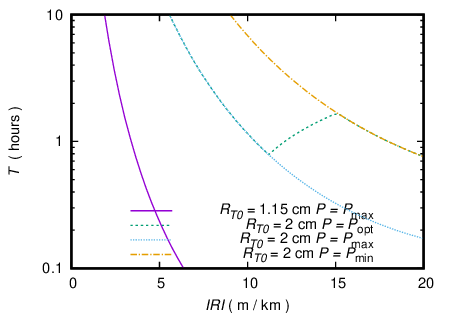}

    (b)
    
   \caption{(a) The root means square acceleration for the
     same cases shown in Figure~\ref{fig_aerodrag}, calculated
     using Equation~\ref{eq_a_rms_b}.  These data
     show that the pressure that minimises the resistance
     force (as in Figure~\ref{fig_resistance_frame}) is not
     always identical with the pressure that minimises
     $a_\mathrm{rms}$.

   (b) The permissible riding time if the Vibration Dosage
   Value is to remain below $150$~m~s$^{-1.75}$, using the data
   in (a) above in conjunction with Equation~\ref{eq_evdv}. \label{fig_a_rms}}
   \end{center}
\end{figure}

Figure~\ref{fig_a_rms} shows the root
mean square acceleration calculated from Equation~\ref{eq_a_rms} for
the conditions of Figure~\ref{fig_aerodrag}.  In conjunction with
Equation~\ref{eq_evdv} these data suggest what others have
surmised
\cite{tarabini_whole-body_2015,roseiro_hand-arm_2016,duc_vibration_2016,doria_experimental-numerical_2021,edwards_thunder_2021},
that cycling activities are likely to breach the thresholds for
VDV exposure.  The data in Figure~\ref{fig_a_rms} tie this conclusion
rather more specifically to road roughness and bicycle characteristics
than previous work but the outcome is much the same.  For instance, if
$VDV=20$~m~s$^{-1.75}$ is the threshold (a typical statutory limit),
this will be breached in 1 minute of cycling with
$a_\mathrm{rms}=10$~m~s$^{-2}$.  If we accepted
$VDV=150$~m~s$^{-1.75}$ (as occurred in the Paris-Roubaix race), then
the threshold will be breached in about 4 hours.  However, whether
thresholds set for protection from occupational exposure over a
working life can be appropriately applied to intense sporting
activities undertaken sporadically is far from obvious, and is
presumably likely to be the subject of further research.  Whether
cycling as a means of daily transport causes dangerous VDV exposure
seems also unknown.

\section{Concluding remarks}
\label{sec_conclusion}

Previous works have established that what is here called roughness
resistance is an important effect, but they have not offered any
systematic understanding or quantification.  This paper shows that
this effect can be described by surprisingly simple expressions that
depend only on two clearly defined and easily measured physical
parameters, namely the vertical stiffness of the bicycle and the
International Roughness Index of the surface on which the bicycle
travels (Equation~\ref{eq_roughness_resistance}).  The stiffness of the
bicycle can be approximately decomposed into a frame stiffness and a
tyre stiffness (Equation~\ref{eq_bicycle_stiffness}).  Either of these can
be dominant.  In traditional high performance bicycles, the tyre
stiffness is much less than the frame stiffness; for a typical modern
bicycle, both are important; and for certain modern bicycles with
exceptionally low frame stiffness, the frame stiffness is the dominant
factor.  There are immediate implications for the optimal choice of
tyre diameter, because increasing the tyre diameter reduces the tyre
stiffness and consequently the roughness resistance.  However,
increasing the tyre diameter also increases the aerodynamic drag area.
Which of these effects is the most important depends on the objectives
of the cyclist and on the circumstances (in particular, on the road
roughness).  There is no generally optimal solution.

The cycling public may find these considerations recondite.
However, practically all cyclists are exposed to vibration due to road
roughness.  As we have seen, there is reason to be concerned about this.
Equations~\ref{eq_evdv} and \ref{eq_a_rms} can be used to connect road
condition surveys with the Vibration Dose Value experienced by
cyclists in general, in a manner that may be of interest to
authorities concerned with both pavement maintenance
and public health.

\section*{Disclosure statement}

No potential conflict of interest was reported by the author.

\bibliographystyle{tfnlm}
\bibliography{better}

\end{document}